\documentstyle[twoside]{article}
\newcount\Mac  \Mac=0  % devo mettere Mac=1 se sto lavorando sul file Mac
\newcommand{\ifMac}[2]{\ifnum\Mac=1 #1 \else #2 \fi}
\def\putps(#1,#2)(#3,#4)#5#6{\ifnum\Mac=1 \put(#1,#2){\special{picture #5}}
\else  \put(#3,#4){\includegraphics{#6}} \fi}
\def\Red  {}
\def\Black{}
\def\Blue {}
\newcommand{\GeV}{\,{\rm GeV}}

\newcommand{\TeV}{\,{\rm TeV}}

\newcommand{\NP}{Nucl. Phys.}

\newcommand{\PRL}{Phys. Rev. Lett.}
\newcommand{\PL}{Phys. Lett.}
\newcommand{\PR}{Phys. Rev.}
\newcommand{\CL}{\,\hbox{\rm C.L.}}
\newcommand{\NO}{\hbox{---}}

\newcommand{\eq}[1]{~(\ref{eq:#1})}
\newcommand{\sys}[1]{~(\ref{sys:#1})}

\def\Ord{{\cal O}}  
 
\def\circa#1{\,\raise.3ex\hbox{$#1$\kern-.75em\lower1ex\hbox{$\sim$}}\,}
\makeatletter
\def\art{\@ifnextchar[{\eart}{\oart}}
\def\eart[#1]#2#3#4#5#6{{\rm #2}, {\em #3 \rm #4} {\rm (#6) #5 ({\em #1})}}
\def\hepart[#1]#2{{\rm #2, \em#1}}
\newcommand{\oart}[5]{{\rm #1}, {\em #2 \rm #3} {\rm (#5) #4}}

\newcounter{alphaequation}[equation]
\def\thealphaequation{\theequation\hbox to
0.6em{\hfil\alph{alphaequation}\hfil}}
\def\eqnsystem#1{
\def\@eqnnum{{\rm (\thealphaequation)}}
\def\@@eqncr{\let\@tempa\relax \ifcase\@eqcnt \def\@tempa{& & &} \or
  \def\@tempa{& &}\or \def\@tempa{&}\fi\@tempa
  \if@eqnsw\@eqnnum\refstepcounter{alphaequation}\fi
\global\@eqnswtrue\global\@eqcnt=0\cr}
\refstepcounter{equation} \let\@currentlabel\theequation \def\@tempb{#1}
\ifx\@tempb\empty\else\label{#1}\fi
\refstepcounter{alphaequation}
\let\@currentlabel\thealphaequation
\global\@eqnswtrue\global\@eqcnt=0 \tabskip\@centering\let\\=\@eqncr
$$\halign to \displaywidth\bgroup \@eqnsel\hskip\@centering
$\displaystyle\tabskip\z@{##}$&\global\@eqcnt\@ne
\hskip2\arraycolsep\hfil${##}$\hfil& \global\@eqcnt\tw@\hskip2\arraycolsep
$\displaystyle\tabskip\z@{##}$\hfil
\tabskip\@centering&\llap{##}\tabskip\z@\cr}
\def\endeqnsystem{\@@eqncr\egroup$$\global\@ignoretrue} \makeatother

\begin{document}
\twocolumn[
\centerline{\bf May.\ 1999 \hfill    IFUP--TH/21--99}
\centerline{\bf hep-ph/9905281 \hfill SNS--PH/99--09} \vspace{1cm}
\centerline{\LARGE\bf\Red What is the limit on the Higgs mass?}

\bigskip\bigskip\Black
\centerline{\large\bf Riccardo Barbieri} \vspace{0.3cm}
\centerline{\em Scuola Normale Superiore and INFN, sezione di Pisa, I-56126 Pisa, Italia}\vspace{0.3cm}
\centerline{\large  and }\vspace{0.3cm}
\centerline{\large\bf Alessandro Strumia}\vspace{0.3cm}
\centerline{\em Dipartimento di fisica, Universit\`a di Pisa and INFN,
sezione di Pisa,  I-56126 Pisa, Italia}

\bigskip\bigskip\Blue

\centerline{\large\bf Abstract}
\begin{quote}\large\indent
We obtain the bounds on all the gauge invariant, flavour symmetric,
CP-even operators of dimension 6 that can affect the electroweak precision tests.
For the preferred Higgs mass of about $100\GeV$, their minimal scales
range from 2 to 10 TeV.
Depending on the individual operator, these limits are often significantly stronger than those
quoted in the literature, when they exist at all.
The impact, if any, of these bounds on the upper limit con the Higgs mass
itself is discussed.
\end{quote}\Black
\vspace{1cm}]

\noindent

\paragraph{1}
What happens of the upper bound on the Higgs mass, $m_h$, set by electroweak precision tests
if physics, although described by the perturbative Standard Mo\-del
at currently explored energies, changes regime at the lowest scale $\Lambda$ compatible with
existing experiments?
An objective answer to this important question is impossible in general since
we do not know what will happen at $\Lambda$, as we do not even know, for the same
reason, the precise value of $\Lambda$ itself.
For sure we only know that the upper bound on $m_h$ of about $250\GeV$ at $95\%\CL$~\cite{mh<250}
holds for `high enough' $\Lambda$.

We cannot be content with such a statement, however.
In fact it is rather obvious to see what should happen in order to evade this bound.
To some extent, the plausibility of this circumstance can also be judged.
A related problem is the estimate of the lowest value of $\Lambda$.
All these issues we intend to discuss in this paper, triggered by a recent
interesting work of Hall and Kolda~\cite{HK}.

The framework we consider is quite general (although not completely general) and standard.
Physics at scales well below $\Lambda$ is described by the effective Lagrangian
\begin{equation}\label{eq:Leff}
{\cal L}_{\rm eff}={\cal L}_{\rm SM}+\sum_i \frac{c_i}{\Lambda^p}{\cal O}_i^{(4+p)}
\end{equation}
where ${\cal L}_{\rm SM}$ is the SM Lagrangian, including the Higgs doublet getting a vev
$\langle{H}\rangle=(0,v)$;
$i$ runs over all the operators ${\cal O}_i^{(4+p)}$ of dimension $4+p$, $p\ge1$,
consistent with the classical symmetries of ${\cal L}_{\rm SM}$ in the limit of vanishing Yukawa couplings
and $c_i$ are unknown dimensionless couplings.
We are not interested here in the origin of ${\cal L}_{\rm eff}$, which should anyhow be a decent
approximation of a generic situation with new physics at $\Lambda$, provided the flavour problem can be kept under
control and no new degree of freedom exists with a mass scale close to the Fermi scale.
Our purpose is to compare as carefully as possible the predictions of\eq{Leff} with the
electroweak precision measurements.

The list of experimental observables~\cite{mh<250} is given in table~1, other than the $Z$ mass $M_Z$ and the
Fermi constant $G_{\rm F}$, both known with great precision\footnote{Further precision measurements that
could be added to this list are the atomic parity violation, the neutrino-nucleon
cross sections and tests of quark-lepton universality.
They are not considered here because their inclusion would not change any of the conclusions.}.

\begin{table*}[t]
$$\begin{array}{rlll}
\Gamma_Z &=&(2.4939\pm0.0024)\GeV & \hbox{total $Z$ width}\\
R_h  &=& 20.765\pm0.026 & \hbox{$\Gamma(Z\to \hbox{hadrons})/\Gamma(Z\to\mu^+\mu^-)$}\\
R_b  &=& 0.21680\pm0.00073 & \hbox{$\Gamma(Z\to b \bar b)/\Gamma(Z\to \hbox{hadrons})$}\\
\sigma_h &=& (41.491\pm0.058)\hbox{nb} & \hbox{$e\bar{e}$ hadronic cross section at $Z$ peak}\\
s^2_{\rm eff}  &=& 0.23157\pm0.00018 & \hbox{effective $\sin^2 \theta_{\rm W}$
(from $\ell\bar{\ell}$ and $b\bar{b}$ asymmetries)}\\
M_W  &=&(80.394\pm0.042)\GeV & \hbox{pole $W$ mass}\\
m_t  &=& (174.3\pm5.1)\GeV & \hbox{pole top masss}\\
\alpha_3(M_Z)  &=& 0.119\pm0.004 & \hbox{strong coupling}\\
\alpha_{\rm em}^{-1}(M_Z)  &=& 128.92\pm0.036 & \hbox{electromagnetic coupling}\\
\end{array}$$
\caption{\em Electroweak precision observables.}
\end{table*}

\begin{table*}[t]
$$\begin{array}{rlll}
\multicolumn{3}{c}{\hbox{\Blue Dimension 6 operators}}&
\qquad\hbox{Effects on precision observables}\Black \\  \hline
\Ord_{WB}&=&(H^\dagger \tau^a H) W^a_{\mu\nu} B_{\mu\nu} & \delta e_3=2/\tan\theta_{\rm W}\\ 
\Ord_{H}&=&|H^\dagger D_\mu H|^2 &\delta e_1=-1\\
\Ord_{LL}&=&\frac{1}{2}(\bar{L}\gamma_\mu \tau^a L)^2 & \delta G_{\rm VB}=2\\ 
\Ord_{HL}' &=&i(H^\dagger D_\mu \tau^a H)(\bar{L}\gamma_\mu \tau^a L) &
\delta g_{Ve}=\delta g_{Ae}=-1\qquad \delta g_{V\nu}=\delta g_{A\nu}=+1,\qquad \delta G_{\rm VB}=4\\
\Ord_{HQ}' &=&i(H^\dagger D_\mu \tau^a H)(\bar{Q}\gamma_\mu \tau^a Q) &
\delta g_{Vd}=\delta g_{Ad}=-1\qquad \delta g_{Vu}=\delta g_{Au}=+1\\
\Ord_{HL} &=&i (H^\dagger D_\mu H)(\bar{L}\gamma_\mu L) &
\delta g_{Ve}=\delta g_{Ae}=-1\qquad \delta g_{V\nu}=\delta g_{A\nu}=-1\\
\Ord_{HQ} &=&i (H^\dagger D_\mu H)(\bar{Q}\gamma_\mu Q) &
\delta g_{Vd}=\delta g_{Ad}=-1\qquad \delta g_{Vu}=\delta g_{Au}=-1\\
\Ord_{HE} &=& i (H^\dagger D_\mu H)(\bar{E}\gamma_\mu E) &\delta g_{Ve}=-\delta g_{Ae}=-1\\
\Ord_{HU} &=& i (H^\dagger D_\mu H)(\bar{U}\gamma_\mu U) &\delta g_{Vu}=-\delta g_{Au}=-1\\
\Ord_{HD} &=& i (H^\dagger D_\mu H)(\bar{D}\gamma_\mu D) &\delta g_{Vd}=-\delta g_{Ad}=-1
%\Ord_{LB} &=& i(\bar{L}\gamma_\mu D_\nu L)B^{\mu\nu} &\delta g_{Ve}=+\delta g_{Ae}=-\frac{1}{2}g_2\tan\theta_{\rm W}\\
%\Ord_{EB} &=& i(\bar{E}\gamma_\mu D_\nu E)B^{\mu\nu} &\delta g_{Ve}=-\delta g_{Ae}=-\frac{1}{2}g_2\tan\theta_{\rm W}\\
%\Ord_{LW} &=& i(\bar{L}\tau^a \gamma_\mu D_\nu L) W^a_{\mu\nu}& \delta g_{Ve}=+\delta g_{Ae}=\frac{1}{2}g_2
\end{array}$$
\caption{\em Dimensions 6 operators affecting the electroweak precision tests,
with their contributions,
up to a common factor $c_i(v/\Lambda)^2$, to the various form factors.
The effect of the hermitian conjugate to the operators from $\Ord_{HL}'$ to
$\Ord_{HD}$ is included.}
\end{table*}

\begin{table*}[t]
$$\begin{array}{c|cc|cc|cc}
\Blue m_h\Black &\multicolumn{2}{|c|}{\Blue100\GeV\Black}&
\multicolumn{2}{|c|}{\Blue300\GeV\Black}&
\multicolumn{2}{|c}{\Blue800\GeV\Black}\\ 
\Blue c_i\Black&\Blue-1\Black&\Blue+1\Black&\Blue-1\Black&\Blue+1\Black&\Blue-1\Black&\Blue+1\Black\\ \hline \hline
\Blue\Ord_{WB} \Black&10 & 9.7&6.9 & \NO&6.0 & \NO \\
\Blue\Ord_{H} \Black&5.5 & 4.5&3.7 & \NO&3.2 & \NO \\
\Blue\Ord_{LL} \Black&8.1 & 5.9&6.3 & \NO& \NO & \NO \\
\Blue\Ord_{HL}'\Black&8.8 & 8.3&6.6 & \NO& \NO & \NO \\
\Blue\Ord_{HQ}'\Black&6.6 & 6.9& \NO & \NO& \NO & \NO \\
\Blue\Ord_{HL}\Black&7.6 & 8.9& \NO & \NO& \NO & \NO \\
\Blue\Ord_{HQ}\Black&5.7 & 3.5& \NO & 3.7& \NO & \NO \\
\Blue\Ord_{HE}\Black&8.8 & 7.2& \NO & 7.1& \NO & \NO \\
\Blue\Ord_{HU}\Black&2.4 & 3.3& \NO & \NO& \NO & \NO \\
\Blue\Ord_{HD}\Black&2.2 & 2.5& \NO & \NO& \NO & \NO 
\end{array}\qquad
\begin{array}{c|cc|cc|cc}
\Blue m_h\Black &\multicolumn{2}{|c|}{\Blue100\GeV\Black}&
\multicolumn{2}{|c|}{\Blue300\GeV\Black}&
\multicolumn{2}{|c}{\Blue800\GeV\Black}\\ 
\Blue c_i\Black&\Blue-1\Black&\Blue+1\Black&\Blue-1\Black&\Blue+1\Black&\Blue-1\Black&\Blue+1\Black\\
\hline\hline
\Blue\Ord_{WB}\Black&8.8 & 8.5&6.4 & 23&5.6 & \NO \\
\Blue\Ord_{H}\Black&4.7 & 4.0&3.4 & 11&2.9 & \NO \\
\Blue\Ord_{LL}\Black&6.8 & 5.4&5.3 & 13& \NO & \NO \\
\Blue\Ord_{HL}'\Black&7.6 & 7.3&6.0 & 18& \NO & \NO \\
\Blue\Ord_{HQ}'\Black&5.7 & 6.0&9.2 & 7.2& \NO & \NO \\
\Blue\Ord_{HL}\Black&6.7 & 7.6&12 & 8.6& \NO & \NO \\
\Blue\Ord_{HQ}\Black&4.7 & 3.2&9.1 & 3.3& \NO & \NO \\
\Blue\Ord_{HE}\Black&7.5 & 6.4&15. & 6.0& \NO & \NO \\
\Blue\Ord_{HU}\Black&2.1 & 2.8&2.8 & 4.1& \NO & \NO \\
\Blue\Ord_{HD}\Black&1.9 & 2.2&2.2 & 3.8& \NO & \NO 
\end{array}$$
\caption{$95\%\CL$ and $99\%\CL$ bounds on $\Lambda/\TeV$ for the single operators for
$c_i=-1$ and $c_i=+1$ and different values of $m_h$.}
\end{table*}

\paragraph{2}
We focus our attention on the operators in\eq{Leff} of the lowest dimension,
$p=2$, which affect the electroweak precision observables $A_i$ in table~1.
A minimal set of such operators is given in table~2.
This set is minimal in the sense that any other operator that contributes to the $A_i$ can be
written as a combination of them, up to operators that give null contribution\footnote{Other than
dependent upon the chosen observables, the set of
operators in table~2 reflects a choice of basis. Innumerable analyses of
the corrections from various sets of operators to some electroweak
precision tests exist in the literature, starting from the early paper~\cite{BW}.
For a recent example, see e.~g.~\cite{Belayev}.
Ref~\cite{GW} might have been the first to give a complete set for
the operators involving bosons only. We have not seen anywhere in the
literature a complete minimal basis for the observables in table 1. It is
easy to write the four bosonic operators in~\cite{GW} as combinations of the
operators in table 2, as shown to us by Riccardo Rattazzi.
The bounds on the $\Lambda$ parameters associated with the operators
$\Ord_1$ and $\Ord_2$, in the notation of~\cite{GW}, range from 2 to 3 TeV.}.
In any operator involving fermion fields, a sum over diagonal flavour indices $f=\{1,2,3\}$ is understood,
as required by the $\hbox{U}(3)^5$ global invariance.
Implicit are also the SU(2) indices $i=\{1,2\}$, on which the Pauli matrices $\tau^a$ act.
For example the operator $\Ord_{LL}$ is
$\frac{1}{2}
(\bar{L}_{f i}\gamma_\mu \tau^a_{ij} L_{f j})
(\bar{L}_{f' i'}\gamma_\mu \tau^a_{i'j'} L_{f' j'})$.
Finally, the addition of a hermitian conjugate term is also left understood whenever needed, which is the case
for all the operators involving both fermions and bosons\footnote{In the operators involving fermions,
the fields have fixed chirality, left for $L$ and $Q$ and right for
$E$, $U$, $D$, and standard normalization of the kinetic terms.}.

The operators of table~2 can affect the electroweak precision observables in table~1 through their contributions to the
following form factors, precisely defined, e.g., in~\cite{BCF}:
\begin{enumerate}
\item vacuum polarization amplitudes $\delta e_1$, $\delta e_3$, related to the $S$, $T$ parameters~\cite{1990};

\item relative contributions to the $\mu$-decay amplitude $\delta G_{\rm VB}$
not coming from vacuum polarization corrections to the $W$-boson, already included in $\delta e_i$;

\item vertex contributions $\delta g_{Vf}$ and $\delta g_{Af}$
to the vector and axial form factors at $q^2=M_Z^2$ of the amplitude
$Z\to f\bar{f}$ for any fermion $f$
\end{enumerate}
As well known, the corrections from $\delta e_1$, $\delta e_3$ and $\delta G_{\rm VB}$ have universal character.
As such, they affect all the electroweak observables
through the parameters $\epsilon_i$, via~\cite{BCF}
\begin{eqnsystem}{sys:eps}
\delta \epsilon_1 &=& \delta e_1- \delta G_{\rm VB}\\
\delta \epsilon_2 &=& -\delta G_{\rm VB}\\
\delta \epsilon_3 &=& \delta e_3
\end{eqnsystem}
in the way explicitly given in~\cite{ABC}.
On the contrary, the vertex corrections are specific of the individual channel.
They correct the widths $Z\to f\bar{f}$ and the asymmetries in an obvious way
from their tree level expressions in terms of $g_{Vf}$ and $g_{Af}$.
Explicit formulas for the corrections to all the electroweak precision observables are as follows\footnote{In
$s^2_{\rm eff}$ one neglects the small effects due to $\delta g_{Vu,d}$ and $\delta g_{Au,d}$.
This approximation is needed to define $s^2_{\rm eff}$ itself.}
\begin{eqnarray}\nonumber
M_W &=& M_W^{\rm SM}
(1+0.72 \delta e_1-0.43 \delta e_3-0.22 \delta G_{\rm VB}) \\ \nonumber
s^2_{\rm eff} &=& s^{2\rm SM}_{\rm eff}
(1-1.43 (\delta e_1- \delta G_{\rm VB})+1.86 \delta e_3+\\&& \nonumber
 -0.163 \delta g_{Ae}+2.16 \delta g_{Ve})\\ \nonumber
\Gamma_Z &=&\Gamma_Z^{\rm SM}
(1+1.35  (\delta e_1- \delta G_{\rm VB})-0.45 \delta e_3+\\ \nonumber
&&-1.23 \delta g_{Ad}+0.82 \delta g_{Au} - 0.85 \delta g_{Vd}+\\&& \nonumber
-0.03 \delta g_{Ve}+0.32 \delta g_{Vu}+\\&&+\nonumber
0.41 (\delta g_{V\nu }+\delta g_{A\nu}-\delta g_{Ae}))\\ \nonumber
R_h &=& R_h^{\rm SM}(1+0.27 (\delta e_1-\delta G_{\rm VB})-0.36 \delta e_3+\\ \nonumber
&&-1.78 \delta g_{Ad}+3.98 \delta g_{Ae}+1.19 \delta g_{Au}+\\&& \nonumber
-1.23 \delta g_{Vd}+0.30 \delta g_{Ve}+0.46 \delta g_{Vu})\\ \nonumber
\sigma_h &=&\sigma_h^{\rm SM}(1-0.03(\delta e_1-\delta G_{\rm VB})+0.03 \delta e_3+\\ \nonumber
&&+0.68 \delta g_{Ad}-3.16  \delta g_{Ae}-0.46 \delta g_{Au}+\\&&- \nonumber
0.82(\delta g_{A\nu} +\delta g_{V\nu})+\\&&+ \nonumber
0.47 \delta g_{Vd}-0.24 \delta g_{Ve}-0.18 \delta g_{Vu})\\ \nonumber
R_b&=&R_b^{\rm SM}(1-0.06( \delta e1-\delta G_{\rm VB})+0.08 \delta e_3+\\&& \nonumber
-0.92 \delta g_{Ad}-1.19 \delta g_{Au}+\\&&
-0.64 \delta g_{Vd}-0.46 \delta g_{Vu})\label{eq:obs}
% Q_W &=& Q_W^{\rm SM}+0.015(\delta e_1-\delta  G_{\rm VB})+1.3 \delta e_3-
% 5.7 \delta g_{Vd}-5.1 \delta g_{Vd}-2\delta g_{Ae})
\end{eqnarray}
where $A_i^{\rm SM}=A_i^{\rm SM}(m_h,m_t,\alpha_{\rm s}(M_Z),\alpha_{\rm em}(M_Z))$ stands for the SM prediction,
dependent on the Higgs and the top masses other than on the strong and electromagnetic gauge couplings.
As to the crucial dependence of the $A_i^{\rm SM}$ on $m_h$, it is worth remembering that it is also
universal (as for $m_t$) occurring through the same $\epsilon$ parameters.
To a good numerical approximation, it is~\cite{ABC}
\begin{eqnsystem}{sys:h}
\delta \epsilon_1(m_h) &\approx & -0.74~10^{-3}~\ln (m_h/M_Z)\\
\delta \epsilon_2(m_h) &\approx & +0.35~10^{-3}~\ln (m_h/M_Z)\\
\delta \epsilon_3(m_h) &\approx & +0.65~10^{-3}~\ln (m_h/M_Z)
\end{eqnsystem}
Finally, to be able to use eq.s\eq{obs} directly, one needs the contributions from the operators
${\cal O}_i$ to the different form factors.
Such contributions are given in the same table~2.

\begin{figure*}[t]
\begin{center}
\begin{picture}(17.7,5)
\putps(-0.5,0)(-0.5,0){f3op}{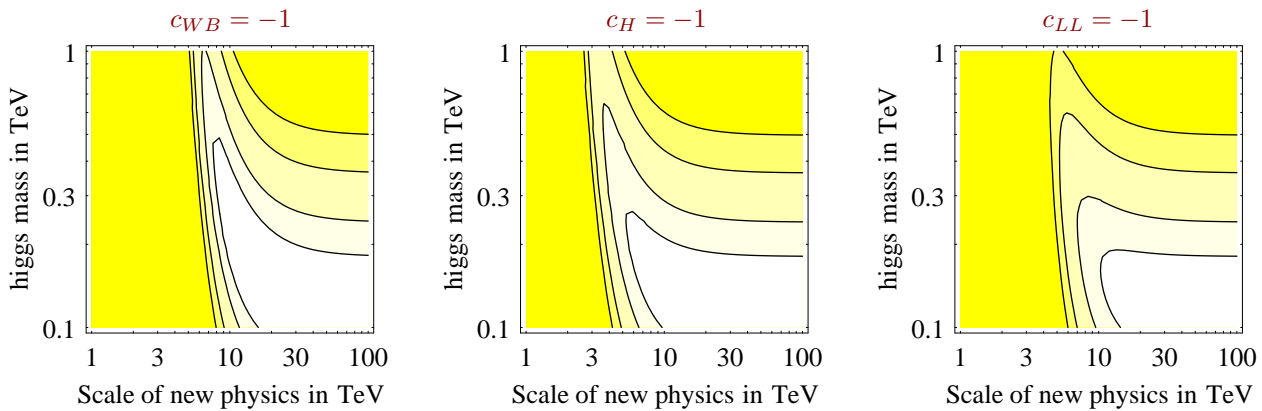}\Red
\put(2.2,5.3){$c_{WB}=-1$}
\put(8.0,5.3){$c_{H}=-1$}
\put(13.8,5.3){$c_{LL}=-1$}\Black
\end{picture}
\caption[SP]{\em Level curves of $\Delta \chi^2=\{1,2.7,6.6,10.8\}$
%that correspond to $\{68\%,90\%,99\%,99.9\%\}\CL$
for the first 3 operators in table~2 ($\Ord_{WB}$, $\Ord_{H}$, and ${\cal O}_{LL}$ in the order)
and $c_i=-1$.
\label{fig:CP1}}
\end{center}\end{figure*}

\begin{figure*}[t]
\begin{center}
\begin{picture}(17.7,5)
\putps(-0.5,0)(-0.5,0){fhisto}{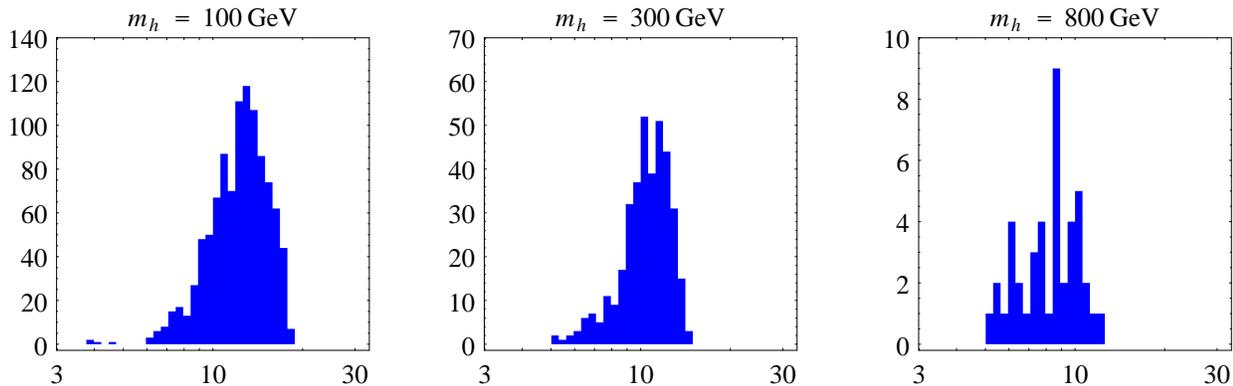}
\end{picture}
\caption[SP]{\em Distribution of the limits on $\Lambda/\TeV$ at $95\%\CL$
for the set of 1024 ``theories'' defined in the text.
\label{fig:fhisto}}
\end{center}\end{figure*}

\paragraph{3}
The first question that we can answer, the only one we can actually answer in a rather objective way, is the limit
set by the electroweak precision tests, as defined above, on the individual operators in table~2 or on the
corresponding $\Lambda$ parameters with $c_i=+1$ or $c_i=-1$, since what counts is the interference
with the SM amplitudes.
This we do by making a fit of the observables in table~1 with variable $m_h$, $m_t$, $\alpha_{\rm s}(M_Z)$,
 $\alpha_{\rm em}(M_Z)$ and $\Lambda$ itself.
The results are shown in table~3 for different values of the Higgs mass $m_h$ and different
confidence levels\footnote{A more conservative error on
$\alpha_{\rm em}(M_Z)$,  $\Delta \alpha_{\rm em}^{-1}(M_Z)=0.08$, does not change these bounds
in a significant way, except for the operator $\Ord_{WB}$, where the limits on $\Lambda$ are reduced by about $1\TeV$.
Adding the measurement of atomic parity violation to the fitted data
would sligtly improve only the weakest bounds in table~3.}.
A blank in table~3 corresponds to the fact that, for the given Higgs mass, no value of
$\Lambda$ allows a $\chi^2< \chi^2_{\rm min}+\Delta \chi^2$ with $\Delta\chi^2=\{3.85,6.6\}$.
For $\chi^2_{\rm min}$ we use its value in the SM fit at $\Lambda=\infty$.
The chosen levels of $\Delta\chi^2$ practically correspond to the conventional $95\%$ and $99\%$ C.L.s.
%for the different C.L.s ($95\%$ and $99\%$)
We have checked that in all cases the minimum is obtained for light Higgs.

Note that these limits are often significantly stron\-ger than those quoted
in the literature.
As an example, the limit recently obtained at the highest
LEPII energies~\cite{LEP} on the $LL$ operator
contributing to $e\bar{e}\to\ell_f\bar{\ell}_f$, in the notation of~\cite{OPAL},
reflects itself in a limit on the scale of the operator $\Ord_{LL}$ in table~2
of about $1.7\TeV$, to be compared with the limits on $\Ord_{LL}$ in table~3.

The results in table~3 have a simple interpretation.
In most cases an increasing Higgs mass makes an acceptable fit increasingly problematic
for whatever value of $\Lambda$ and overall sign: the addition of the corresponding
individual operator cannot cure the difficulty of the pure SM fit for high Higgs masses.
In a few cases only, a fit with a Higgs mass definitely higher than $300\GeV$ is still possible.
Which operators (and which signs for their coefficients)
make this possible is clear by comparing eq.s\sys{eps} with eq.s\sys{h}
in view of table~2: they are the operators that affect the parameters $\epsilon_1$ and $\epsilon_3$,
which are the most sensitive to the changes of the Higgs mass.
Also $\epsilon_2$ changes with the Higgs mass, but to a lesser extent.
Furthermore its change can be compensated by a contribution to $\delta G_{\rm VB}$ that, however,
also affects $\epsilon_1$ with a relative wrong sign.
This is why the operators that contribute to $\delta G_{\rm VB}$ can improve the high $m_h$ fit but only
to a relatively marginal extent.

Quite clearly, the limits in the last columns of table~3 correspond more properly to a range
of values for the related $\Lambda$ parameters~\cite{HK}, since in absence of the correction from the
additional operator no fit would be possible at all.
In fig.~\ref{fig:CP1} we show the contour plot of $\Delta \chi^2$ in the plane
$(\Lambda,m_h)$ for the first 3 operators in table~2, that can lead, for an appropriate choice of the sign
(negative in all cases), to a dilution of the bound on the Higgs mass.

\paragraph{4}
An important question arises,
although difficult to answer in an objective way: how plausible is it that a motivated theory
exists which gives rise to either of the operators
$\Ord_H$ or $\Ord_{WB}$, or both, with the appropriate signs, and suppresses, at the same time,
all other operators, or at least the most disturbing of them?
We remind the reader that in the literature around 1990 several ad hoc examples were exhibited
in renormalizable field theories~\cite{Georgi},
although generally involving the exchange of non decoupled new particles.
It should be possible to construct examples~\cite{Casalbuoni}
using particles genuinly decoupling
in their heavy mass limit.
They would serve as existence proofs.
Whether one can call these examples `motivated' is a debatable matter.

A possible way to illustrate the problem in a numerical manner is the following.
We can consider the set of ``theories'', each defined by the Lagrangian\eq{Leff} with all
$|c_i|=1$ but with a specific choice of the signs of the 10 coefficients in front of the 10 operators
of table~2: they are $2^{10}=1024$ different ``theories''.
For each of them one can ask what is the limit set on $\Lambda$ by the electroweak
precision tests for different values of the Higgs mass, as we did for the individual operators in the
previous section. The result is shown in fig.~\ref{fig:fhisto} in the form of a histogram for
$m_h=\{100,300,800\}\GeV$.
Notice the decrease in the number of ``theories'' that can give an acceptable fit for increasing $m_h$:
from 1024 for $m_h=100\GeV$ to 367
for $m_h=300\GeV$ and 
43 for $m_h=800\GeV$.
Notice also, as expected, the lowering of the mass $\Lambda$ for the few ``theories'' that can
support a heavy Higgs.

\begin{figure}[t]
\begin{center}
\begin{picture}(6,6)
\putps(-0.5,0)(-0.5,0){fav}{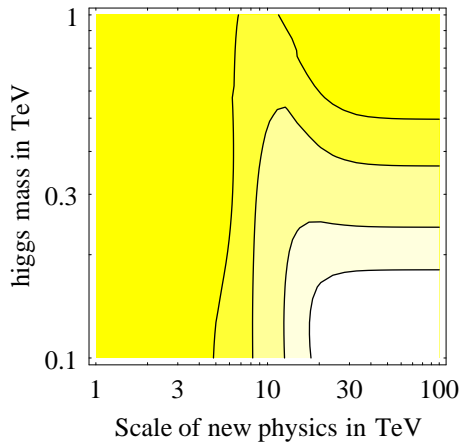}
%\put(2.7,5.3){\ref{fig:chiq}a}
%\put(8.5,5.3){\ref{fig:chiq}b}
%\put(14.3,5.3){\ref{fig:chiq}c}
\end{picture}
\caption[SP]{\em Contour plot of the `average' $\chi^2_{\rm av}$, at
$\Delta \chi^2=\{1,2.7,6.6,10.8\}$.
%that correspond to $\{68\%,90\%,99\%,99.9\%\}\CL$
\label{fig:fmean}}
\end{center}\end{figure}

Finally in fig~\ref{fig:fmean} we dare to show in the plane $(\Lambda, m_h)$ the contour plot of
an average $\chi^2_{\rm av}$ probability, defined according the laws of probability as
$$\exp(-\chi^2_{\rm av}/2)=\sum_{c=1}^{1024} \exp(-\chi^2_c/2)$$
where $c$ runs over all possible ``theories''.
Notice that, if this were the plot relevant for the ``true'' theory, which clearly is not,
the electroweak precision tests alone would give a limit on the scale $\Lambda$ of about $10\TeV$.

\paragraph{5} In conclusion, we have obtained the bounds on all the flavour symmetric and CP-conserving
dimension six operators from the electroweak precision tests available today.
For the preferred Higgs mass of about $100\GeV$ they range from $2$ to $10\TeV$ for the corresponding
$\Lambda$ parameters.

What is the impact of these bounds, if any, on the upper limit on the Higgs mass?
A conservative point of view would be the following.
These limits make it unlikely that there be any new physics below $10\TeV$,
so that the limits on the Higgs mass can be safely studied in the pure SM.
Here one is barring the existence of new degrees of freedom well below $\Lambda$,
as one forgets the problem of the unnaturally light Higgs,
about 100 times lower than the minimal scale of new physics.

Alternatively one can suppose~\cite{HK} that one or two appropriate operators exist that dominate over the others
and, at the same time, more or less precisely counteract the effect on the $\chi^2$ of a heavy Higgs with a
mass well above the limit set by the SM analysis.
Not likely, we believe, but possible.
In such case the problem of Higgs naturalness could be somewhat alleviated,
since the Higgs mass would be closer to the necessarily low scale of new physics.

One should not forget, of course, the possibility that there be new degrees of freedom so close to
the mass of the SM particles that integrating them out in an effective Lagrangian like\eq{Leff} does
not make sense at all to descrive physics below a TeV, in the putative range of the Higgs mass.
These new particles, if they exist, have been so far elusive.
LEP, {\sc Tevatron} and finally LHC will settle the issue.

\paragraph{Acknowledgments}
We are grateful to Riccardo Rattazzi for many useful discussions.

\end{document}